\documentclass[pra, reprint, showpacs, nofootinbib]{revtex4-1}
\usepackage{graphicx}
\usepackage{amsthm, amsmath, amssymb}
\usepackage[fonts]{faust}
\usepackage{dirac}

\usepackage[colorlinks=true,linkcolor=blue]{hyperref}
\usepackage{cleveref}

\graphicspath{ {./images/} }

\newcommand\defn[1]{{\bfseries\slshape #1\/}}
\newcommand\hc{{\mathrm{hc}}}

\newcommand\mc{{\mathrm{mc}}}
\newcommand\pa{{\mathrm{pa}}}

\begin{document}

\title{On the paraxial approximation in quantum optics II:\\ 
Henochromatic modes of a Maxwell field}

\author{M.\ Fernanda \surname{Jongewaard de Boer} and Christopher Beetle}

\date{January 15, 2023}

\begin{abstract}
A companion paper \cite{Scalarpaper} has argued that the best way to associate single-particle quantum states of a scalar field to the modes of a narrowly collimated beam of classical radiation modeled in the paraxial approximation uses the ``henochromatic'' states previously introduced by Sudarshan, Simon and Mukunda \cite{SSMs, SSMv}.  This paper extends that result to Maxwell fields, again emphasizing the central role of unitarity in defining the association. 
 The principal new technical element in the present discussion has to do with the intertwining of polarization and spatial degrees of freedom in the resulting single-photon states.
\end{abstract}

\maketitle

\section{Introduction}

In a companion paper \cite{Scalarpaper} we detailed how the ``henochromatic'' fields proposed by Sudarshan, Simon and Mukunda \cite{SSMs,SSMv} offer a uniquely preferred way, among a broad class of similar proposals \cite{DG, AW, CPB, ACM, AML, AVNW}, to associate exact, single-particle, quantum states of a scalar field to the modes of a (linearly polarized) laser beam as they are understood in the paraxial approximation.  Most importantly, the mapping from paraxial waves to henochromatic states is \emph{unitary}, allowing the logic of the resulting quantum states to mirror that of the underlying paraxial wave modes.  The present paper extends our previous analysis, and its proof of the unique advantages of henochromatic states, to the case of a Maxwell field.

There are many well-known ways to quantize a Maxwell field.  We will focus on the Coulomb quantization because, in addition to its simplicity (for pure radiation fields), the paraxial approximation for Maxwell fields is itself best understood by first imposing the Coulomb gauge condition classically.  The incorporation of this gauge condition, however, introduces the main technical difficulty in our analysis of Maxwell fields that did not arise for scalar fields.  Namely, in passing to the paraxial approximation, it is conventional to simplify the Coulomb condition $\vec\grad \cdot \vec A(\vec r, t) = 0$ to the transversality condition $\vec k_0 \cdot \vec A(\vec r, t) = 0$, where $\vec k_0$ is the principal wave vector around which the support of the paraxial field $\vec A(\vec r, t)$ is clustered in Fourier space.  The simplification is considerable in the paraxial approximation as it serves to ``disentangle'' the polarization and spatial degrees of freedom of the field.  But our goal is to associate the resulting ``disentangled'' fields to genuine states in the single-particle sector of the Hilbert space of the Coulomb quantization, which definitely remain ``entangled'' by the exact Coulomb gauge condition.  We show below, however, that this difficulty is indeed merely technical, and that henochromatic Maxwell fields still provide a uniquely preferred association of the type we seek.

This paper is organized as follows.  Section 2 reviews the paraxial approximation for Maxwell fields, paying special attention to the role of the Coulomb gauge condition.  Section 3 recalls the standard quantization of the Maxwell field in the Coulomb gauge, emphasizing an approach that avoids focusing undue attention on a fixed basis of (plane-wave) states in defining the Fock structure of the resulting Hilbert space.  Section 4 outlines several desirable features a prospective mapping from wave modes in the paraxial approximation to exact, single-photon states ought to have, and then proceeds to show that the mapping defined by the henochromatic states is unique, at least within a large class of similar mappings, in exhibiting all of these features.  We conclude with some comments in Section 5.

\section{The Paraxial Approximation for Maxwell Fields} \label{M:CPA}

A pure radiation field in Maxwell theory satisfies the homogeneous Maxwell equations 
\begin{equation}
\begin{alignedat}{2}
    \vec\grad \cdot \vec E(\vec r, t) &= 0 &\qquad
    \vec\grad \times \vec E(\vec r, t) &= - \pd{\vec B}{t}(\vec r, t) \\[1ex]
    \vec\grad \cdot \vec B(\vec r, t) &= 0 &\qquad
    \vec\grad \times \vec B(\vec r, t) &= \frac{1}{c^2} \pd{\vec E}{t}(\vec r, t).
\end{alignedat}
\end{equation}
Any pair of electric and magnetic fields, $\vec E(\vec r, t)$ and $\vec B(\vec r, t)$, satisfying these equations can be derived from a Coulomb vector potential of the form 
\begin{equation}\label{CG:Adef}
    \vec A(\vec r, t) := \vec\grad \times \int_{\Re^3}\frac{\vec B(\vec\xi, t)}{4 \pi \|\vec r - \vec\xi\|}\, \ed^3 \xi, 
\end{equation}
provided they are sufficiently well-behaved asymptotically.  More precisely, restricting the integral here to a finite volume $\Sigma \subset \Re^3$, one can show mathematically that 
\begin{align}
    \vec B(\vec r, t)
        &= \vec\grad \times \vec A(\vec r, t)
            + \vec\grad \oint_{\delta\Sigma} \frac{\vec{\hat n}(\vec \xi) \cdot \vec B(\vec\xi, t)}{4 \pi \|\vec r - \vec\xi\|}\, \ed^2 \xi
\intertext{and}
    \vec E(\vec r, t) 
        &= - \pd{\vec A}{t}(\vec r, t) 
            + \vec\grad \oint_{\delta\Sigma} \frac{\vec{\hat n}(\vec\xi) \cdot \vec E(\vec\xi, t)}{4 \pi \|\vec r - \vec\xi\|}\, \ed^2 \xi
        \notag\\&\hspace{5em}
            - \vec\grad \times \oint_{\delta\Sigma}\frac{\vec{\hat n}(\vec\xi) \times \vec E(\vec\xi, t)}{4 \pi \|\vec r - \vec\xi\|}\, \ed^2 \xi 
\end{align}
for all $\vec r \in \Sigma$.  A field is ``well-behaved asymptotically'' if the surface integrals here vanish as the boundary $\delta\Sigma$ is taken to infinity.

There is an easier way to implement this condition, however, at least for fields that can be written as well-defined superpositions of plane waves.  Namely, the Coulomb-gauge potential for each individual plane-wave mode $\vec B(\vec r, t) = \vec B_0\, \ee^{\ii (\vec k \cdot \vec r - c \|\vec k\| t)}$ comprising such a field is 
\begin{equation}
    \vec A(\vec r, t) = \lim_{\epsilon\to 0^+} \vec\grad \times \frac{\vec B_0\, \ee^{\ii (\vec k \cdot \vec r - c \|\vec k\| t)}}{\|\vec k\|^2 + \epsilon^2},
\end{equation}
where the regulating factor $\epsilon$ accounts for the distributional nature of the Fourier transform of the Coulomb potential.  The limit $\epsilon \to 0^+$ here exists for any $\vec k \ne \vec 0$, and any radiation field with no static (\textit{i.e.}, $\vec k = \vec 0$) component can therefore be written uniquely in terms of a well-defined Coulomb-gauge potential given by \cref{CG:Adef}.  Indeed, these potentials are manifestly in one-to-one correspondence with such radiation fields, confirming explicitly that the Coulomb condition \emph{completely} fixes the gauge of $\vec A(\vec r, t)$.

The paraxial approximation for Maxwell fields can be motivated in much the same way that it is for scalar fields \cite{Scalarpaper}.  To summarize, a general (positive-frequency) superposition of plane-wave Maxwell modes has the form 
\begin{equation} \label{pwExp}
	\vec A^{\!+}(\vec r, t) = \int d^{3} k \; \rho(\vec k) \, \vec{\mathcal{A}}(\vec k) \frac{\ee^{\ii (\vec k \cdot {\bf r} - c \|\vec k\| t )}}{\sqrt{(2 \pi)^{3} \, 2 \| \vec k\| \, \rho(\vec k)}},
\end{equation}
where $\vec{\mathcal{A}}(\vec k)$ is generally a complex, vector-valued function of $\vec k$ and $\rho(\vec k) > 0$ is a real, scalar-valued function corresponding to the density of the plane-wave states being superposed\relax
%%%
\footnote{As noted in the scalar case, the choices $\rho(\vec k) = 1$, corresponding to a uniform density of states in the Fourier 3-space of the inertial frame implicitly chosen by the time coordinate $t$ in \cref{pwExp}, and $\rho(\vec k) = 1 / 2 \|\vec k\|$, corresponding to a uniform density of states on the forward light cone in Fourier 4-space, are the most common choices.  The former leads to simpler commutation relations at the quantum level, while the latter has better relativistic covariance properties.  But in principle one could choose \emph{any} $\rho(\vec k) > 0$.}\relax
%%%
.  Restricting attention to amplitude profiles $\vec{\mathcal{A}}(\vec k)$ satisfying the \defn{tranversality condition} 
\begin{equation} \label{Trans}
	 \vec k \cdot  \vec{\mathcal{A}} (\vec k) = 0
\end{equation}
for each $\vec k$ implements the Coulomb gauge condition for \cref{pwExp}.  The paraxial approximation then applies when the support of $\vec{\mathcal{A}}(\vec k)$ is (mostly) restricted to a small region around a given $\vec k_0 \ne \vec 0$ in Fourier space.  More precisely, we demand that $\vec{\mathcal{A}}(\vec k) \approx 0$ unless $\|\vec k - \vec k_0\| \ll k_0 := \|\vec k_0\|$.  As in the scalar case, we refer to the set of radiation fields having the form of \cref{pwExp}, satisfying the transverality condition of \cref{Trans}, and obeying this loose condition on the support of $\vec{\mathcal{A}}(\vec k)$ as the \defn{paraxial regime} of Maxwell theory.

In our nomenclature, fields in the paraxial \emph{regime} \cite{AW, AVNW, AML, CPB, ACM, SSMs, SSMv} are exact, Coulomb-gauge solutions of the (positive-frequency) wave equation, which are approximately monochromatic as a natural consequence of the localization of $\vec{\mathcal{A}}(\vec k)$ in Fourier space.  In contrast, fields in the paraxial \emph{approximation} are exactly monochromatic, but typically satisfy both the wave equation and the Coulomb gauge condition only approximately.  These approximating fields emerge by first restricting \cref{pwExp} to be \emph{exactly} monochromatic, \textit{i.e.}, by choosing $\vec{\mathcal{A}}(\vec k)$ to have distributional support on the sphere of radius $k_0$ about the origin in Fourier space.  Aligning the $+z$-axis along the principal wave vector $\vec k_0$, and using the transverse wave vector $\vec q := \vec k = \vec{\hat z} \vec{\hat z} \cdot \vec k$ as coordinates on the (forward hemi-)sphere in Fourier space, yields 
\begin{equation}\label{mcfld}
    \vec A\!^\mc(\vec s, z, t) = \int \frac{\ed^2 q}{2 \pi}\, \vec{\mathcal{F}}^\mc(\vec q)\, \ee^{\ii \vec q \cdot \vec s + \ii z \sqrt{k_0^2 - \|\vec q\|^2} - \ii c k_0 t}, 
\end{equation}
where the 2-dimensional vector $\vec s$ denotes the transverse coordinates in the $xy$-plane and 
\begin{equation}
    \vec{\mathcal{A}}(\vec k) = \sqrt{\frac{4 \pi k_0}{\rho(\vec k)}}\, \delta \biggl( k_z - \sqrt{k_0^2 - \|\vec q\|^2} \biggr)\, \vec{\mathcal{F}}^\mc(\vec q)
\end{equation}
in \cref{pwExp}.  The transversality condition of \cref{Trans} demands that 
\begin{equation}\label{mcCoul}
    \mathcal{F}^\mc_z(\vec q) 
        = - \frac{\vec q \cdot \vec{\mathcal{F}}^\mc(\vec q)}{\sqrt{k_0^2 - \|\vec q\|^2}} 
\end{equation}
for such a monochromatic field.

The paraxial approximation for Maxwell fields is most accurate when modeling monochromatic fields in the paraxial regime highlighted above.  It works by replacing the square roots \cref{mcfld,mcCoul} as follows.  First, one expands the square root in the second exponential factor from \cref{mcfld} in a Taylor series 
\begin{equation}\label{kzExp}
    \sqrt{k_0^2 - \|\vec q\|^2} = k_0 - \frac{\|\vec q\|^2}{2 k_0} + \cdots, 
\end{equation}
and then drops the higher-order terms not shown here.  The retained quadratic term captures the physical phenomenon of the diffractive spreading of a narrowly collimated beam as it propagates along its longitudinal axis.  Second, one replaces the right side of \cref{mcCoul}, which already is of sub-leading order $\|\vec q\| / k_0$, with zero.  The resulting vector potential then has no component along the longitudinal ($+z$-)axis of the beam, and its remaining components in the transverse ($xy$-)plane reflect the two-dimensional space of polarization states of the Maxwell field.  Replacing the carrier frequency $ck_0 \mapsto ck$ for notational simplicity in what follows, the corresponding spacetime fields have the form \cite{AW,ACM,AVNW}
\begin{subequations}\label{parEnv}
\begin{align}\label{parMod}
    \vec A^{\!\pa}(\vec s, z, t) ={}&{} \vec\Xi(\vec s, z)\, \ee^{\ii k (z - ct)} 
\intertext{with}\label{parFour}
    \vec\Xi(\vec s, z) :={}&{} \int \frac{\ed^2 q}{2 \pi}\, \vec{\mathcal{F}}(\vec q)\, \ee^{\ii \vec q \cdot \vec s}\, \ee^{- \ii \frac{\|\vec q\|^2}{2 k} z}, 
\end{align}
\end{subequations}
where $\mathcal{F}_z(\vec q)$, and thus $\Xi_z(\vec s, z)$, vanishes.  Although the envelope function $\vec\Xi(\vec s, z)$ here is now vector-valued, it satisfies the same \defn{paraxial wave equation} 
\begin{align}\label{parEq}
    0 ={}&{} \biggl( 2 \ii k\, \pdby{z} + \triangle \biggr) \vec\Xi(\vec s, z) 
        \notag\\[1ex]
        :={}&{} \biggl( 2 \ii k\, \pdby{z} + \pdby[2]{x} + \pdby[2]{y} \biggr) \vec\Xi(\vec s, z) 
\end{align}
as in the scalar theory.  Furthermore, again as in the scalar case, the space of such envelope functions admits a Schr\"odinger-like inner product 
\begin{align}\label{parIP}
    \eprod{\vec\Xi_1}{\vec\Xi_2} 
        &= \int \ed^2 s\, \vec{\bar{\Xi}}_1(\vec s, z) \cdot \vec\Xi_2(\vec s, z) 
            \notag\\[1ex]
        &= \int \ed^2 q\, \vec{\bar{\mathcal{F}}}\!_1(\vec q) \cdot \vec{\mathcal{F}}\!_2(\vec q), 
\end{align}
where the first integral has the same value for every cross-section ($z = \text{const.}$) of the beam.  This inner product gives the space $\mfs{K}_k$ of fields from \cref{parEnv} sharing a common carrier frequency $ck$ the structure of a Hilbert space.  

Note that \cref{parEq,parIP} can both be recast in terms of the non-zero components $\Xi_{x,y}(\vec s, z)$ of the envelope function.  This reduces the paraxial approximation in Maxwell theory to a pair of independent \emph{scalar} paraxial approximations, one for each (linear) polarization of the beam.  However, this decoupling is quite distinct from what happens even for strictly monochromatic fields that solve the wave equation exactly.  Indeed, one can regard \cref{mcCoul} as a condition \emph{intertwining} the ``polarization'' degrees of freedom of a monochromatic field, loosely associated with the direction of the vector $\vec{A}\!^\mc(\vec s, z, t)$ or $\vec{\mathcal{F}}^\mc(\vec q)$ at a point, with its ``spatial'' degrees of freedom, associated with how the field varies from one point to another.  Simplifying this condition to $\vec A^{\!\pa}_z(\vec s, z, t) = 0$ or $\mathcal{F}\!_z(\vec q) = 0$ in \cref{parEnv} has the effect of disentangling those degrees of freedom.  But the \emph{absence} of such entanglement for Maxwell fields in the paraxial approximation that is the exception, not the rule.  We must pay careful attention below to the question of how Maxwell fields in the paraxial approximation should be ``reentangled'' when associating them to single-particle states of the quantum field.

\section{Quantum Maxwell Theory in the Coulomb Gauge}

% The paraxial approximation for Maxwell fields outlined in the previous section incorporates the Coulomb gauge condition in determining the fields.  For our project of connecting the paraxial approximation to the quantum theory of a Maxwell field, it therefore makes sense to incorporate the Coulomb gauge condition in the quantization.  That is, it is natural to use the Coulomb quantization of the Maxwell field.  This section reviews that quantization process.

The Hilbert space of the Coulomb quantization of Maxwell theory is a Fock space constructed from the space of those positive-frequency solutions 
\begin{equation}
    	\vec A^{\!+}(\vec r , t) 
        = \int \ed^3 k\, \rho(\vec k)\, \vec{\mathcal{A}} (\vec k)\, 
            \frac{\ee^{\ii (\vec k \cdot \vec r - c \|\vec k\| t)}}{\sqrt{(2 \pi)^3\, 2 \|\vec k\|\, \rho(\vec k)}} 
\end{equation}
of the wave equation that also satisfy the Coulomb gauge condition 
\begin{equation}
    \vec\grad \cdot \vec A^{\!+}(\vec r, t)  = 0 
    \qquad\text{or}\qquad 
    \vec k \cdot \vec{\mathcal{A}}(\vec k) = 0.
\end{equation}
Specifically, the single-particle Hilbert space $\mfs{H}$ is the completion of the space of such classical fields in the Hermitian inner product 
\begin{align}
    \iprod[\big]{\vec A^{\!+}_1}{\vec A^{\!+}_2} 
        :={}&{} \frac{\ii}{\hbar c^2} \int \ed^3 r\, \biggl( 
            \vec{\bar A}^{\!+}_1(\vec r, t)\, \cdot \pd{\vec A^{\!+}_2}{t}(\vec r, t) 
        \notag\\&\hspace{6em} 
            - \pd{\vec{\bar A}^{\!+}_1}{t} (\vec r, t)\, \cdot \vec A^{\!+}_2(\vec r, t) \biggr) 
            \notag\\[1ex]
        ={}&{} \frac{1}{\hbar c} \int \ed^3 k\, \rho(\vec k)\, \vec{\bar{ \mathcal{A}}}_1(\vec k) \cdot \vec{  \mathcal{A}}_2(\vec k).
\end{align}
The Fock construction of the multi-particle  Hilbert space $\mfs{F\! H}$ for the quantum Maxwell field theory defines a creation operator $\hat a^\dagger[\vec A^{\!+}]$,  one for each (normalizable) field $\ket{\vec A^{\!+}} \in \mfs{H}$ in the single-particle Hilbert space, and its adjoint annihilation operator $\hat a[\vec{\bar A}^{\!+}]$, which is most naturally labeled by the adjoint vector $\bra{\vec A^{\!+}} \in \mfs{H}^*$ lying in the dual of the single-particle Hilbert space.  These operators satisfy the canonical commutation relations 
\begin{equation}\label{aCCR}
    \comm[\big]{\hat a[\vec{\bar A}{^{\!+}_1}]}{\hat a^\dagger[\vec A^{\!+}_2]} 
        = \iprod[\big]{\vec A^{\!+}_1}{\vec A^{\!+}_2}\, \hat 1 
\end{equation}
by definition, where the inner product on the right is that of the single-particle Hilbert space $\mfs{H}$.

The preceding account of the Coulomb quantization of the Maxwell field is complete and entirely equivalent to its conventional, textbook construction.  The key benefit of this approach for our purposes is that it avoids emphasizing the plane-wave basis 
\begin{equation}\label{pwMode}
    \vec\Phi_{\vec k, \alpha}(\vec r, t) 
        := \sqrt{\frac{\hbar c}{2 \|\vec k\| \rho(\vec k)}}\, 
            \frac{\ee^{\ii (\vec k \cdot \vec r - c \|\vec k\| t)}}{(2 \pi)^{3/2}}\, \vec\varepsilon_\alpha (\vec k), 
\end{equation}
for the Hilbert space $\mfs{H}$ of single-particle quantum states of the quantum field, where $\vec\varepsilon_{1,2}(\vec k)$ is a fixed, but arbitrary (and generally complex), orthonormal basis for the 2-dimensional subspace of $\mbb{E}^3$ that is perpendicular to the argument $\vec k$.  Indeed, our goal is to associate a single-particle, quantum state to \emph{any} given field in the paraxial approximation.  We do not fix a basis for the Hilbert space $\mfs{K}_k$ of paraxial Maxwell waves defined in he previous section.  Even if we did, however, the quantum states in $\mfs{H}$ corresponding to those basis states in $\mfs{K}_k$ would bear no particular relation to a basis fixed \textit{a priori} in $\mfs{H}$.  Introducing such an \emph{a priori} structure can only serve to obscure the correspondence we aim to establish.  We therefore prefer the basis-independent approach \cite{DG, MS, D} outlined above.  The main disadvantage of our approach is that one cannot write the field operator $\hat{\vec A}(\vec r, t)$ explicitly, as quantum field theory texts conventionally do in terms of the annihilation operators
\begin{equation}
    \hat a_{\vec k, \alpha} := \hat a \bigl[ \vec\Phi_{\vec k, \alpha} \bigr]
\end{equation}
for the priviledged, plane-wave basis states (and their adjoints).  However, the \emph{same} field operator can be defined implicitly via the condition 
\begin{equation}
    \comm[\big]{\hat{\vec A}(\vec r, t)}{\hat a^\dagger[\vec A^{\!+}]} 
        := \vec A^{\!+}(\vec r, t)\, \hat 1 
\end{equation}
in our approach, where $\ket{\vec A^{\!+}} \in \mfs{H}$ is arbitrary.

\section{Paraxial Single-Photon States}

As in the case of scalar fields \cite{Scalarpaper}, our goal now is to construct a mapping $\vec\Xi \mapsto \vec A_{\vec\Xi}$ from the Hilbert space $\mfs{K}_k$ of solutions to the paraxial wave equation (for a given carrier frequency $c k$) to the Hilbert space $\mfs{H}$ of single-particle states in quantum Maxwell theory.  The mapping we seek should have the following properties: 
\begin{enumerate}\renewcommand\theenumi{\Alph{enumi}}
\item It should be \emph{linear}, so that superpositions of the resulting single-particle states exactly mirror those of the underlying paraxial waves.
\item It should be \emph{unitary}, at least in the sense that $\iprod{\vec A_{\vec\Xi}}{\vec A_{\vec\Xi'}}$ vanishes whenever $\eprod{\vec\Xi}{\vec\Xi'}$ does.  This will ensure that the algebra of projection operators associated with the filtering and measurement of single-particle quantum states also mirrors that of the underlying paraxial waves.
\item It should be \emph{consistent} with our results for the scalar model in the following sense.  Enforcing the Coulomb gauge condition on $\vec A_{\vec\Xi}(\vec r, t)$ generally has the effect of intertwining the spatial degrees of freedom of a single-photon state with its polarization, as discussed at the end of \cref{M:CPA} above.  However, there are some paraxial waves $\vec\Xi(\vec s, z)$ for which the approximating field of \cref{parEnv} \emph{already} satisfies the Coulomb gauge condition.  We assert that no such intertwining should be required for those states, and that the mapping $\vec\Xi \mapsto \vec A_{\vec\Xi}$ should then be dictated by the scalar-field mapping of \cite{Scalarpaper}.
\item It should be \emph{covariant} in the sense that rotating $\vec\Xi(\vec s, z)$ about the optical ($+z$-)axis, or rigidly translating it in Euclidean space, induces the same transformation of $\vec A_{\vec\Xi}(\vec r, t)$ relative to the inertial frame in which we enforce the Coulomb gauge condition.  
\item It should be \emph{scale invariant} in the sense that the definition of $\vec A_{\vec\Xi}(\vec r, t)$ in terms of $\vec\Xi(\vec s, z)$ should not privilege any particular length scale other than that set by the carrier frequency $ck$.
\end{enumerate}
In addition to these requirements, and again as in the scalar model, we will restrict our attention to mappings of the general form 
\begin{equation}\label{sbPsi}
    \vec A_{\vec \Xi} (\vec s, z, t) 
        = \int \frac{\ed^2 q}{2 \pi}\, \vec{\mathcal{F}'}(\vec q)\, \ee^{\ii \vec q \cdot \vec s}\, \ee^{\ii \kappa(\vec q, k) z}\, \ee^{-\ii \omega(\vec q, k) t}.
\end{equation}
The functions $\kappa(\vec q, k)$ and $\omega(\vec q, k)$ remain arbitrary for the moment.  All examples of such a mapping we are aware of in the literature \cite{DG, AW, AVNW, CPB, AML, ACM, SSMv} have this general form, though the choices of these two functions vary.  Unlike the scalar model, however, note that $\vec{\mathcal{F}'}(\vec q)$ in \cref{sbPsi} is not necessarily equal to $\vec{\mathcal{F}}(\vec q)$ in \cref{parEnv}.  This is because of the need to intertwine spatial and polarization degrees of freedom for some Maxwell fields mentioned above, a need which does not arise for scalar fields.

Before we explore the consequences of the conditions (A--E) outlined above, recall that $\vec{A}_{\vec\Xi}(\vec s, z, t)$ in \cref{sbPsi} should belong to the single-particle Hilbert space $\mfs{H}$ of the quantum Maxwell theory.  That is, it should satisfy both the (positive-frequency) wave equation and the Coulomb gauge condition, whence 
\begin{align}\label{fWEq}
    \omega(\vec q, k) &= c \sqrt{\kappa^2(\vec q, k) + \|\vec q\|^2} 
\intertext{and}\label{fCgc}
    0 &= \bigl( \vec q + \kappa(\vec q, k)\, \vec{\hat z} \bigr) \cdot \vec{\mathcal{F}'}(\vec q).
\end{align}
Equivalently, the latter condition requires that the Cartesian components of $\vec{\mathcal{F}'}(\vec q)$ along the $z$-axis and along the axis parallel to $\vec q$ within the $xy$-plane satisfy 
\begin{equation}\label{Fzcomp}
    \mathcal{F}'_z(\vec q) = - \frac{\|\vec q\|}{\kappa(\vec q, k)}\, \mathcal{F}'_\shortparallel(\vec q).
\end{equation}
The remaining component of $\vec{\mathcal{F}'}(\vec q)$, the one along the axis perpendicular to $\vec q$ within the $xy$-plane, is unconstrained by this condition.

Now we proceed to the conditions (A--E) laid out above.  Because the Fourier transform is a linear operation, the linearity condition (A) will hold automatically provided that the mapping $\vec{\mathcal{F}}(\vec q) \mapsto \vec{\mathcal{F}'}(\vec q)$ is linear.  \Cref{Fzcomp} constrains the allowed linear mappings, but does not determine one uniquely.

The consistency condition (C) further constrains the linear mapping $\vec{\mathcal{F}}(\vec q) \mapsto \vec{\mathcal{F}'}(\vec q)$ as follows.  The paraxial field of \cref{parEnv} satisfies the Coulomb gauge condition if and only if $\vec q \cdot \vec{\mathcal{F}}(\vec q) = 0$.  In this case, condition (C) demands that the mapping of Maxwell fields we seek should mimic that of the scalar fields we analyzed previously \cite{Scalarpaper}.  The role analogous to $\vec{\mathcal{F}'}(\vec q)$ in the scalar case, however, was played simply by the analogue of $\vec{\mathcal{F}}(\vec q)$ itself.  Thus, we condition (C) demands that 
\begin{equation}\label{conF}
    \vec{\mathcal{F}'}(\vec q) = \vec{\mathcal{F}}(\vec q) 
    \quad\text{whenever}\quad 
    \vec q \cdot \vec{\mathcal{F}}(\vec q) = 0.
\end{equation}
But the mapping $\vec{\mathcal{F}}(\vec q) \mapsto \vec{\mathcal{F}'}(\vec q)$ must be linear, so this relationship generalizes to 
\begin{equation}\label{Focomp}
    \mathcal{F}'_{\!\shortperp}(\vec q) := \mathcal{F}^{}_{\!\shortperp}(\vec q) 
\end{equation}
for an \emph{arbitrary} paraxial field, where $\mathcal{F}^{}_{\!\shortperp}(\vec q)$ denotes the component of $\vec{\mathcal{F}}(\vec q)$ perpendicular to $\vec q$ in the $xy$-plane, \textit{i.e.}, the component unconstrained by \cref{Fzcomp}.

Next we turn to the unitarity condition (B).  As in the scalar case, we simply compute the (relativistic) inner product of the single-particle states $\vec{A}_{\vec{\Xi}_{1, 2}}(\vec s, z, t)$ associated by \cref{sbPsi} to an arbitrary pair of paraxial waves $\vec{\Xi}_{1, 2}(\vec s, z)$, potentially with different carrier frequencies $c k_{1,2}$.  We find 
\begin{widetext}
\begin{align}
    \iprod{\vec{A}_{\vec\Xi_1}}{\vec{A}_{\vec\Xi_2}} 
        :={}&{} \frac{4\pi}{\hbar c^2}\, \delta(k_2 - k_1)
            \int \ed^2 q\, 
                \frac{\omega(\vec q, k_1)}{\bigl| \pd{\kappa}{k}(\vec q, k_1) \bigr|}\, 
                \vec{\bar{\mathcal{F}}'_1}(\vec q) 
                \cdot 
                \vec{\mathcal{F}'_2}(\vec q).
\end{align}
\end{widetext}
The integral here is proportional to the (non-relativistic) inner product of \cref{parIP} for all choices of $\vec\Xi_{1,2}(\vec s, z)$ if and only if there exists a function $\Omega(k)$ such that 
\begin{equation}\label{unitarity}
    \frac{\omega(\vec q, k)}{\bigl| \pd{\kappa}{k}(\vec q, k) \bigr|}\, 
    \vec{\bar{\mathcal{F}}'_1}(\vec q) 
    \cdot 
    \vec{\mathcal{F}'_2}(\vec q) 
    = \Omega(k)\,  
    \vec{\bar{\mathcal{F}}_1}(\vec q) 
    \cdot 
    \vec{\mathcal{F}_2}(\vec q) 
\end{equation}
for all $\vec{\mathcal{F}_{1,2}}(\vec q)$.  But, in the special case where $\vec{\mathcal{F}_{1,2}}(\vec q)$ are both orthogonal to $\vec q$, the inner products on either side are equal by \cref{conF}.  It follows that 
\begin{equation}\label{ucon}
    \frac{\omega(\vec q, k)}{\bigl| \pd{\kappa}{k}(\vec q, k) \bigr|} = \Omega(k) 
\end{equation}
generally.  This is the same relation we found in the scalar case \cite{Scalarpaper}, where we showed that \cref{fWEq,ucon}, together with the covariance condition (D) and the scale invariance condition (E), fixed the unique choices 
\begin{equation}\label{hcfns}
    \frac{\omega(\vec q, k)}{c} = k + \frac{\|\vec q\|^2}{4 k} 
    \quad\text{and}\quad 
    \kappa(\vec q, k) = k - \frac{\|\vec q\|^2}{4 k}
\end{equation}
of the undetermined functions in \cref{sbPsi}, along with $\Omega(k) = ck$.  The same conclusion holds here.

Since \cref{ucon} equates the scalar factors on either side of \cref{unitarity}, it follows from the latter that the linear mapping $\vec{\mathcal{F}}(\vec q) \mapsto \vec{\mathcal{F}'}(\vec q)$ must preserve inner products.  That is, it is a rotation.  \Cref{Focomp} fixes the rotation axis to be that perpendicular to $\vec q$ in the $xy$-plane.  \Cref{Fzcomp} fixes the rotation angle.  Combining all of these results yields 
\begin{widetext}
\begin{equation}\label{hcPsi}
    \vec A^\hc_{\vec \Xi} (\vec s, z, t) 
        = \int \frac{\ed^2 q}{2 \pi}\, \biggl( \vec{\mathcal{F}}(\vec q) 
            - 2\, \frac{\vec{q} + 2 k \vec{\hat z}}{\|\vec q\|^2 + 4 k^2}\, 
                \vec{q} \cdot \vec{\mathcal{F}}(\vec q) \biggr)\, 
                \ee^{\ii \vec q \cdot \vec s}\, 
                \ee^{\ii \bigl( k - \frac{\|\vec q\|^2}{4k} \bigr) z}\, 
                \ee^{- \ii c \bigl( k + \frac{\|\vec q\|^2}{4k} \bigr) t}.
\end{equation}
This is the unique mapping from Maxwell fields in the paraxial approximation to single-photon states in quantum field theory that satisfies the conditions (A--E) above.  As in the scalar case, the single-particle quantum states are henochromatic.

In the scalar case, we also established that the set of all henochromatic fields from \cref{hcPsi}, where (the scalar analogue of) $\vec\Xi(\vec s, z)$ ranges over all possible solutions of the paraxial wave equation with all possible carrier frequencies $ck$, is \emph{complete} in the single-particle Hilbert space.  The analogous result also holds in the vector case.  To see this, expand an arbitrary positive-frequency Maxwell field $\vec A\!^+(\vec r, t)$ satisfying the Coulomb gauge condition in the plane-wave basis of \cref{pwExp}.  Using the resulting amplitude profile $\vec{\mathcal{A}}(\vec q, k_z)$, set 
\begin{equation}\label{FfromA}
    \vec{\mathcal{F}}(\vec q; k) 
        := \sqrt{\frac{4 k^2 + \|\vec q\|^2}{16 \pi k^3}\,  
            \rho \biggl( \vec q, k - \frac{\|\vec q\|^2}{4 k} \biggr)}\, 
            \Biggl[ \vec{\mathcal{A}} \biggl( \vec q, k - \frac{\|\vec q\|^2}{4 k} \biggr) 
                - \biggl( \frac{\vec q}{2 k} + \vec{\hat z} \biggr)\, \mathcal{A}_z \biggl( \vec q, k - \frac{\|\vec q\|^2}{4 k} \biggr) \Biggr] 
\end{equation}
\end{widetext}
Replacing $\vec{\mathcal{F}}(\vec q)$ in \cref{hcPsi} with this expression, and  integrating the resulting henochromatic fields over all carrier frequencies $k > 0$, then reproduces the original $\vec A\!^+(\vec s, z, t)$.  This shows explicitly that every positive-frequency Maxwell field can be written (uniquely) as a superposition of the henochromatic fields from \cref{hcPsi}.  Note that $\vec{\mathcal{A}}(\vec k)$ in \cref{pwExp} depends on the density of states $\rho(\vec k)$, but that the factor of $\rho(\vec k)$ under the square root in \cref{FfromA} renders $\vec{\mathcal{F}}(\vec q; k)$ independent of that choice.

\section{Conclusions}

The principal goal of this paper was to extend to the case of a Maxwell field our previous work \cite{Scalarpaper} demonstrating the uniquley unitary character of the mapping from scalar solutions of the paraxial wave equation to henochromatic single-particle quantum states of a scalar field.  We have shown that indeed it does extend, though some additional care is needed in specifying the ``polarization degrees of freedom'' for the resulting single-particle state.  Once again, as in the scalar case, henochromatic fields of the form we study here have appeared previously in the literature \cite{ACM, SSMv}.  What is new in our discussion is the emphasis on the unitary character of the mapping.  Physically, this mathematical feature of the mapping ensures that all superpositions and projections of henochromatic single-photon states \emph{exactly} mirror those of the underlying beam modes in the paraxial approximation.

\end{document}